\documentclass[12pt,a4paper]{article}
\usepackage[russian,english]{babel}
\usepackage[cp1251]{inputenc}
\usepackage[T2A]{fontenc}
\usepackage{latexsym}
\usepackage{amssymb}
\usepackage{amsmath}
\usepackage{graphicx}

\catcode`@=11
\renewcommand{\section}{\@startsection{section}{1}{0pt}{\medskipamount}
{\medskipamount}{\large\bf}} \numberwithin{equation}{section}
\catcode`@=12

\newcommand{\be}{\begin{equation}}
\newcommand{\ee}{\end{equation}}
\newcommand{\p}[1]{(\ref{#1})}
\def\a{\alpha}

\def\b{\beta}
\def\p{\partial}

\def\t{\theta}
\def\tr{{\rm tr}}
\def\tr{{\rm tr}\,}
\def\Tr{{\rm Tr}\,}
\def\cN{{\cal N}}
\def\cD{{\cal D}}
\def\cA{{\cal A}}

\def\bea{\begin{eqnarray}}
\def\eea{\end{eqnarray}}
\def\nn{\nonumber}

\def\cN{{\cal N}}

\def\f{\frac}

\def\tr{{\rm tr}\,}

\def\nn{\nonumber}

\def\s{\sigma}

\def\d{\delta}
\def\q{\quad}
\def\g{\gamma}

\def\ve{\varepsilon}

\def\O{\Omega}
\def\U{\Upsilon}
\def\sB{\stackrel{\frown}{\square}}

\def\r{\rho}

\def\cW{\cal W}

\def\eq{\eqref}

\numberwithin{equation}{section}

\date{\it  }
\begin{document}

\begin{center}
\vspace{1cm} {\Large\bf Induced low-energy effective action in the
$6D$, $\cN=(1,0)$ hypermultiplet theory on the vector multiplet
background \vspace{1.2cm} }

 {\bf
 I.L. Buchbinder\footnote{joseph@tspu.edu.ru }$^{\,a,b}$,
 B.S. Merzlikin\footnote{merzlikin@tspu.edu.ru}$^{\,a,c}$,
 \boxed{\rm \bf N.G. Pletnev}$^{\,d,e}$
 }

 {\it $^a$ Department of Theoretical Physics, Tomsk State Pedagogical
 University,\\ 634061, Tomsk,  Russia \\ \vskip 0.15cm
 $^b$ National Research Tomsk State University, 634050, Tomsk, Russia \\ \vskip 0.1cm
 $^c$ Department of Higher Mathematics and Mathematical Physics,\\
\it Tomsk Polytechnic University, 634050, Tomsk, Russia\\\vskip
0.1cm $^d$ Department of Theoretical Physics, National Research
Novosibirsk State University,\\  Novosibirsk, 630090 Russia \\
 $^e$ Sobolev Institute of Mathematics, Novosibirsk, 630090 Russia

}
\end{center}

\begin{abstract}

We consider the six dimensional $\cN=(1,0)$ hypermultiplet model
coupled to an external field of the Abelian vector multiplet in
harmonic superspace approach.  Using the superfield proper-time
technique we find the divergent part of the effective action  and
derive the complete finite induced low-energy superfield effective
action. This effective action depends on external field and contains
in bosonic sector all the powers of the constant Maxwell field
strength. The obtained result can be treated as the $6D$,
$\cN=(1,0)$ supersymmetric Heisenberg-Euler type effective action.

\end{abstract}

\section{Introduction}
The study of the various classical and quantum aspects of $6D$
supersymmetric gauge theories is one of the most interesting and
attractive subjects of modern supersymmetric field theory. Such
theories have profound links with M-branes, they admit the
superfield description, in particular, the $\cN=(1,0)$ theories can
be formulated in terms of unconstrained harmonic superfields, the
$\cN=(1,0)$ and $\cN=(1,1)$ theories possess the interesting
ultraviolet behaviour, in the framework of $6D$ super Yang-Mills
theories one can construct the new supersymmetric model called the
tensor hierarchy (see e.g. the recent papers \cite{SSWW},
\cite{IB13}, \cite{Schwarz}, \cite{CDY}, \cite{BP15}, \cite{BP15-1},
\cite{BIS} and the references therein).

In this letter we consider the hypermultiplet model coupled to
external Abelian vector multiplet, where all the fields formulated
in $6D,\,\cN=(1,0)$ harmonic superspace, and derive the complete
induced low-energy superfield effective action. Such effective
action is gauge invariant and depends on space-time constant
superfield strengths of $6D$ vector multiplet. As a result we will
obtain the new Heisenberg-Euler type superfield effective action.

A computation of the effective action is based on superfield
proper-time technique which is a power tool for analysis of the
effective actions in the supersymmetric gauge theories (see the
applications of this technique in the various superfield models e.g.
in \cite{BK}, \cite{BUCH}, \cite{KUZ}). In the case under
consideration the superfield proper-time technique allows us to
preserve the manifest $6D,\, \cN=(1,0)$ supersymmetry and gauge
invariance on all steps of computations and derive the closed form
of the complete low-energy effective action. Calculation of $6D,\,
\cN=(1,0)$ supersymmetric effective action has some similarity with
one in four-dimensional ${\cal N}=2$ theory however, the
six-dimensional theory possesses many specific features which should
be taken into account at calculations.

The letter is organized as follows. Section 2 is devoted to the main
notions of $6D, \cN=(1,0)$ harmonic superspace, including the
formulations of hypermultiplet and vector multiplet in such a
superspace. In section 3 we study a quantum theory of the $6D,\,
\cN=(1,0)$ hypermultiplet coupled to $6D,\, \cN=(1,0)$ external
Abelian vector multiplet and discuss a definition of induced
superfield effective action of the vector multiplet. Section 4 is
devoted to calculations of the effective action. It is shown in
subsection 4.1 that the effective action is on-shell finite, its
divergent part vanishes when the vector multiplet superfield
satisfies the classical equation of motion. It is an essential
property of $6D,\, \cN=(1,0)$ supersymmetry. For comparison, the
corresponding effective actions in $4D$ supersymmetric theories
contain divergences. In subsection 4.2 we consider the calculations
of the complete low-energy superfield effective action and obtain
the final result (\ref{G2}), (\ref{xi}). Although the computational
technique has some familiar aspects with one in $4D$ supersymmetric
theories, the concrete computations in six dimensional
supersymmetric theory contain many specific details. The effective
action (\ref{G2}), (\ref{xi}) can be expanded in power series in
superfield strengths and their spinor derivatives allows us to
construct the $6D,\, \cN=(1,0)$ on-shell superfield invariants of
the vector multiplet. In conclusion we summarize the results
obtained.

We want to emphasize that the problems of the effective action in
the $6D,\, \cN=(1,0)$ hypermultiplet theory on the vector multiplet
background can be considered as the novel type of the external field
problems in quantum field theory.

\section{Basic $6D$ supersymmetric models in $\cN=(1,0)$ harmonic superspace}
The harmonic  $6D$, $\cN=(1,0)$ superspace was introduced in
\cite{Z86}, \cite{HSW},  \cite{ISZ}. It is parameterized by the
central basis coordinates $(x^m, \t^{\a i}, u^{\pm i})$, where
harmonics $u^{\pm i}$ $(\widetilde{u^\pm_i}=u^{\pm i}, u^{+
i}u^-_i=1, (i=1,2))$ belong to the coset R-symmetry of the group
$SU(2)/U(1).$ Also one can introduce the analytical basis
($\zeta^M_A=\{x^m_A, \t^{+ \a}\}, u^\pm_i, \t^{-\a} $) by the rule
 \be
 x^a_A=x^a +i\t^-\g^a\t^+,\q \t^{\pm \a}=u^\pm_i\t^{\a i}.
 \ee
It is notes that the coordinates $(\zeta^M_A, u^\pm_i)$ form a
subspace closed under the $\cN=(1,0)$ supersymmetry transformations.
The covariant harmonic derivatives form  the Lie algebra of $SU(2)$
group ($ [D^{++}, D^{--}]=D^0$) and  in the analytic basis read
 \bea
 D^{\pm\pm}&=&u^{\pm i}\p^{\mp}_i + i\t^{\pm\a} (\g^m \p_m)_{\a\b}\t^{\pm\b} + \t^{\pm\a}\p^\mp_\a
 \,, \nn \\
 D^0 &=& u^{+i}\p^+_i - u^{-i}\p^-_i  + \t^{+\a}\p^+_\a
 -\t^{-\a}\p^-_\a\,,
 \eea
where we have denoted $\p^\pm_i = \f{\p}{\p u^{\pm i}}$ and
$\p^\pm_\a = \f{\p}{\p u^{\pm \a}}$. Analytic subspace allows us to
define the analytical superfields, which satisfy the condition of
the Grassmann analyticity $D^+_\a \phi=0$. The spinor derivatives
$D^{\pm}_\a$ in the analytic basis have the form
 \be
D^+_\a=\f{\p}{\p\t^{-\a}},\q
D^-_\a=-\f{\p}{\p\t^{+\a}}-2i\p_{\a\b}\t^{-\b}, \q
\{D^+_\a,D^-_\b\}=2i\p_{\a\b}. \label{salg}
 \ee
In what follows we also use the following conventions
 \bea
 && (D^\pm)^4 = -\f{1}{4!}\ve^{\a\b\rho\g}D^\pm_\a D^\pm_\b D^\pm_\rho
D^\pm_\g\,, \quad(D^+)^4(\t^-)^4 = 1\,,  \\
&&   (D^+)^{3\a} = -\f16\ve^{\a\b\g\d}D^+_\b D^+_\g D^+_\d\,, \qquad
(D^+)^{3\a}(\t^-)^3_\b = \d_\b^\a \,,
 \eea
and the other notations from the work \cite{BP15-1}.

The simplest basic $6D,\, \cN=(1,0)$ supersymmetric models are ones
of hypermultiplet and vector multiplet.

The $6D,\, \cN=(1,0)$ hypermultiplet is described by the superfields
$q^i(x,\t)\,, i=1,2\,,$ and their conjugate
$\bar{q}_i(x,\t)=(q^i)^\dag$,  under the constraint
 \be
 \label{10q}D_\a^{(i}q^{j)}(x,\t)=0~.
 \ee
The superfield $q^i(x,\t)$ has a short expansion
$q^i(z)=f^i(x)+\t^{\a i}\psi_\a(x) +\ldots$, with a doublet of
massless scalars $f^i$ and the spinor $\psi_\a$ fields. Thus, the
$6D,\, \cN=(1,0)$ hypermultiplet has 4 bosonic and 4 fermionic real
degrees of freedom.

One can use the  analytic superfields in harmonic superspace to
construct the off-shell Lagrangian formulation of the $6D,\,
\cN=(1,0)$ hypermultiplet.  In this case the $6D,\, \cN=(1,0)$
hypermultiplet is described by an unconstrained analytic superfield
$q_A^+(\zeta, u)$
 \be
 D^+_\a q_A^+(\zeta,u)=0~.
 \ee
The analytic superfield $q_A^+(\zeta, u)$ satisfies the reality
condition $\widetilde{(q^{+A})}\equiv q^+_A=\ve_{AB}q^{+B}$, where
the Pauli-G\"ursey index $A=1,2$ is a lowered and raised by the
matrices $\ve_{AB},\ \ve^{AB}.$

The classical model of the $6D,\, \cN=(1,0)$ hypermultiplet is
described by the action
  \be
  \label{actHyper}
  S_q=-\f12\int d\zeta^{(-4)}du \ {q}^{+A} D^{++}q^{+}_A~,
 \ee
where $d\zeta^{(-4)}=d^6xd^{4}\theta^{+}$ is the analytic superspace
integration measure. The corresponding equations of motion follows
from the action \eq{actHyper} and have the form
 \be
 D^{++}q^{+}(\zeta, u)=0~.
 \ee
In principle the formulation above allows us to write down the most
general hypermultiplet self-couplings in the form of the arbitrary
potential ${\cal L}^{(+4)}(q^+, \tilde{q}^+)$ \cite{gios}.

The off-shell $6D,\, \cN=(1,0)$  vector multiplet is realized in
conventional superspace in the following way\footnote{See the
details in \cite{Z86}, \cite{HSW}, \cite{z} \cite{ISZ}.}. First of
all one can introduce the gauge-covariant derivatives $
\cD_M=D_M+\cA_M$, where the flat derivatives $D_M = (D_a, D^i_\a)$
obeying the anti-commutation relations (\ref{salg}) and the
superfields $\cA_M$ are the gauge connection taking the values in
the Lie algebra of the gauge group. The gauge-covariant derivatives
satisfy the algebra
 \bea
&& \{\cD_\a^i,\cD^j_\b\}=-2i\ve^{ij}\cD_{\a\b},\quad
[\cD_\g^i,\cD_{\a\b}]=-2i\ve_{\a\b\g\d}W^{i\d}, \label{alg1}\\
&&[\cD_a, \cD_b] = F_{ab}\,, \label{alg2}
 \eea
where $W^{i\a}$ is the superfield strength  of the  anti-Hermitian
superfield gauge potential. In this paper we consider the
interaction of the hypermultiplet with background Abelian vector
multiplet.

One can solve the constraints \eqref{alg1} and \eqref{alg2} in the
framework of the harmonic superspace. The integrability condition
$\{\cD^+_\a,\cD^+_\b\}=0$ allows us to express the spinor covariant
derivatives in the form $\cD^+_\a=e^{-ib}D^+_\a e^{ib}$, where
$b(z,u)$  is a some Lie-algebra valued harmonic superfield of zero
harmonic U(1) charge. In the $\lambda$-frame, the spinor covariant
derivatives $\cD^+_\a$ coincide with the flat ones,
$\cD^+_\a=D^+_\a=\f{\p}{\p\t^{-\a}}$. In this case the harmonic
covariant derivatives acquire the connection $V^{++}$,
 \be\label{cD++}
 \cD^{++}=D^{++}+V^{++}~,
 \ee
which is an unconstrained analytic potential of the theory. The
component expansion of $V^{++}(\zeta, u)$ in the Wess-Zumino gauge
 \be
V^{++}_{WZ}=\t^{+\a}\t^{+\b}A_{\a\b}(x_A)+(\t^+)^3_\a\lambda^{-\a}(x_A)+3(\t^+)^4Y^{--}(x_A)~,
 \ee
involves  the  physical fields $A_{\a\b} = (\g^m)_{\a\b} A_m$ and
$\lambda^{i\, \a}$ and the auxiliary ones which are collected in the
superfield $Y^{--}$.

Let us introduce the non-analytic harmonic connection $V^{--}(z,
u)$. The superfield $V^{--}(z, u)$ is determined in terms of
$V^{++}$ uniquely as a solution of the zero-curvature condition
\cite{gios}, which in the Abelian case is reduced to
 \be
 \label{zeroc}
 D^{++}V^{--}-D^{--}V^{++}=0.
 \ee
The gauge transformations of the connection $V^{--}$ has the form
$\d V^{--}=-\cD^{--}\Lambda$, where the gauge parameter $\Lambda$ is
an analytic anti-Hermitian superfield. The decomposition of the
superfield $V^{--}$ in terms of the component fields \cite{BP15}
reads
 \bea
 \label{potSYM}
 V^{--} = \t^{-\a}\t^{-\b}v_{\a\b}(x_A,
  \t^+)+(\t^-)^3_\a v^{+\a}(x_A, \t^+)+(\t^-)^4v^{++}(x_A, \t^+).
 \eea
The components of superfields $v_{\a\b}\,, v^{+\a} $ and $v^{++}$
are discussed in details in the \cite{BP15}. For our aims it is
useful to write down only the components of $v^{++}$
 \bea
 && v^{++}=Y^{++}+\t^{+\a}\chi^{+}_\a+\t^{+\a}\t^{+\b}\O_{\a\b}+(\t^+)^3_\a\r^{-\a}+(\t^+)^4\pi^{(-2)},
 \\
&& \chi^+_\a=i\cD_{\a\b}\lambda^{+\b},\q \ve^{\a\b\g\d}\O_{\g\d}
 =-2i\cD^{\a\b}Y^{+-} - 2D^{[\a\g}F^{\b]}_\g-\f{1}{4}\{\lambda^{+[\a},\lambda^{-\b]}\} \,,\nn \\
&&\r^{-\a}=2i\cD^{\a\b}\chi^-_\b\,, \quad
\pi^{(-2)}=\cD^{\a\b}\cD_{\a\b}Y^{--}
-\f12\{\lambda^{-\a},\cD_{\a\b}\lambda^{-\b}\}\,.  \nn
 \eea
With the help of the connection $V^{--}$ we construct the spinor and
the vector superfield connections ${\cal A}^-_\a=-D^+_\a V^{--},\q
{\cal A}_{\a\b}=\f{i}{2}D^+_\a D^+_\b V^{--}$ and determine the
field strength (in the $\lambda$-frame)
 \be
 \label{defW}W_\lambda^{+\a}=-\f14 (D^+)^{3\a} V^{--},\ee
The Bianchi identities lead to relations
 \be
 D^+_\a W^{-\a}=D^-_\a W^{+\a}, \q D_\a^\pm
 F_{ab}=iD_{[a}(\g_{b]})_{\a\b}W^{\pm\b}.
 \ee
The vector superfield strength, $F_\a^{\b}=(\g_{ab})_\a^{\b}F^{ab}$,
is defined as follows
 \bea
 F_\a^\b=(D^-_\a W^{+\b}-D^+_\a W^{-\b}) = 2 N_\a^\b. \label{N}
 \eea
The other useful consequences of the Bianchi identities are
 \bea
 && D^+_\b W^{+\a}=\f14\d^\a_\b Y^{++}\,, \quad  Y^{++} = -(D^+)^4V^{--}\,, \quad
  D^{++} Y^{++}=0, \label{identity0}  \\
 && \q W^{-\a}=D^{--}W^{+\a}\,, \quad D^{--}Y^{++} = 2 D^-_\a
 W^{+\a}\,, \quad  \label{identity} D^{++}W^{+\alpha}=0\,.
 \eea
These relations define the superfield $Y^{++}$ which will be used
further.

The superfield action of $6D,\,\cN=(1,0)$ SYM theory has the form
\cite{Z86,HSW,z,ISZ}
 \be
 \label{actZ}S_{SYM}=\f{1}{f^2}\sum_{n=1}^\infty\f{(-1)^{n+1}}{n}\tr\int
 d^{14}z du_1\ldots du_n\f{V^{++}(z,u_1)\ldots
 V^{++}(z,u_n)}{(u^+_1u^+_2) \ldots (u^+_{n}u^+_1)},
 \ee
where $f$ is the dimensional coupling constant ($[f]=-1$). The
corresponding equations of motion read
 \be
 Y^{++}=(D^+)^4V^{--}=0. \label{eqY}
 \ee
The connection between component fields of $W^{+\a}$ and $V^{++}$ is
caused by the zero-curvature condition (\ref{zeroc}) and the
definition (\ref{defW}).

\section{Superfield effective action}

The effective action $\Gamma[V^{++}]$, induced by hypermultiplet
matter, is defined by
 \be\label{EAhyper}
 e^{i\Gamma[V^{++}]}=\int\cD q^+\cD \tilde{q}^+
\exp\Big(-i\int d\zeta^{(-4)}\tilde{q}^+\cD^{++}q^+\Big)~.
 \ee
The expression(\ref{EAhyper}) yields
 \be
 \label{Tr}\Gamma[V^{++}]=i\Tr\ln\cD^{++}=-i\Tr\ln G^{(1,1)}.
 \ee
Here $G^{(1,1)}(\zeta_1,u_1|\zeta_2,u_2) =
\langle\tilde{q}^{+}(\zeta_1,u_1){q}^{+}(\zeta_2,u_2)\rangle$ is the
superfield Green function in the $\tau$-frame. This Green function
is analytic with respect to both arguments and satisfies the
equation
 \be\label{eqG}
 \cD_1^{++}G_\tau^{(1,1)}(1|2)=\d_A^{(3,1)}(1|2)~.
 \ee
Here $\d_A^{(3,1)}(1|2)$ is the appropriate covariantly analytic
delta-function
 \be
\d_A^{(q,4-q)}=(D^+_2)^4\d^{14}(z_1-z_2)\d^{(q,-q)}(u_1,u_2)\,.
 \ee
Like in four-dimensional case \cite{BUCH} and \cite{KUZ}  we will
act by the operator $(\cD_1^{--})^2$  on both sides of (\ref{eqG})
 \bea
 \cD^{++}_1(\cD_1^{--})^2 G^{(1,1)}(1|2)=(\cD_1^{--})^2\d_A^{(3,1)}(1|2)= 2\cD^{++}_1
(\cD^+_2)^4\f{\d^{14}(z_1-z_2)}{(u^+_1u^+_2)^3}~.
 \eea
Now, since the equation $D^{++}f^{-|q|} = 0$ has only the trivial
solution $f^{-|q|} = 0$, after the action of the operator
$(\cD^+_1)^4$ we obtain:
 \bea
(\cD^+_1)^4(\cD^{--}_1)^2G^{(1,1)}(1|2) = -8\sB
G^{(1,1)}(1|2)=2(\cD^+_1)^4(\cD^+_2)^4\f{\d^{14}(z_1-z_2)}
{(u^+_1u^+_2)^3}~.
 \eea
Thus we obtain
 \be
\label{GREEN}
 \sB G_\tau^{(1,1)}(1|2)=-\f{1}{4}(\cD^+_1)^4(\cD^+_2)^4\f{\d^{14}(z_1-z_2)}{(u^+_1u^+_2)^3},
 \ee
where $1/(u^+_1u^+_2)^3$  is a special harmonic distribution. In Eq.
(\ref{GREEN}) the operator $\sB$ is the covariantly analytic
d’Alembertian ($[\cD^+_\a,\sB]=0$) which arises when
$(\cD^+)^4(\cD^{--})^2$  acts on the analytical superfield and has
the form
 \be\label{sBox}
 \sB=-\f18(\cD^+)^4(\cD^{--})^2|=\cD_a\cD^a+W^{+\a}\cD^-_\a
+Y^{++}\cD^{--}-\f{1}{4}(D^-_\a W^{+\a}).
 \ee

The operator $\sB$ acts on the space of covariantly analytic
superfields. Let us introduce a new second-order operator $\Delta$,
 \be\label{sDelta}
 \Delta = \sB - W^{-\a}\cD^+_\a\,,
 \ee
which coincides with $\sB$ on the space of covariantly analytic
superfields\footnote{Note that in $6D,\, \cN=(1,0)$ hypermultiplet
theory, the operator (\ref{sDelta}) differs from the analogical
operator in $4D, {\cal N}=2$ hypermultiplet theory}. We have to note
that the Green function, $G^{(1,1)}(1|2)$ , is analytic with respect
to both arguments thus we obtain
 \be \label{Gr2}
 G^{(1,1)}(1|2)=-\f{1}{4
 \Delta}(\cD^+_1)^4(\cD^+_2)^4\f{\d^{14}(z_1-z_2)}{(u^+_1u^+_2)^3}.
 \ee
Like in four- and five-dimensional cases \cite{BUCH}, \cite{KUZ} one
can obtain the useful identity
 \be\label{Delta}
(\cD^+_1)^4(\cD^+_2)^4\f{\d^{14}(z_1-z_2)}{(u^+_1u^+_2)^3}=(\cD^+_1)^4
\Big\{(u^+_1u^+_2)(\cD^-_1)^4 -(u^-_1u^+_2)\Omega^{--} -4\sB
\f{(u^-_1u^+_2)^2}{(u^+_1u^+_2)}\Big\} \d^{14}(z_1-z_2)\,.
 \ee
Here we have introduced the notation
 \bea
 \Omega^{--} = i\cD^{\a\b}\cD^-_\a\cD^-_\b + 4W^{-\a}\cD^-_\a - (D^-_\a
W^{-\a})~. \label{O}
 \eea
This identity is used later for computing the effective action.

The  definition (\ref{Tr}) of the one-loop effective action is
purely formal. The actual evaluation of the effective action can be
done in various ways (see e.g. \cite{BUCH}, \cite{KUZ}). Further we
will follow \cite{KUZ} and use the relation
 \bea
 \Gamma[V^{++}]&=& \Gamma_{g=0}+\int_0^1 dg\p_g\Gamma(g V)=-i \int_0^1
 dg \Tr(V^{++}G^{(1,1)}(gV)) \nn \\
 &=& -i\int d\zeta_{1}^{(-4)} du_1 V^{++}\,\int_0^1
 dg\,G^{(1,1)}(1|2)|_{2=1}\,, \label{Gamma0}
 \eea
where $G^{(1,1)}(gV)$ means the Green function depending on the
superfield $gV^{++}$.

\section{Calculation of the effective action}
Let us discuss a generic scheme of the calculations. We substitute
the Green function \eqref{Gr2} to the effective action \eq{Gamma0}
and one-loop effective action takes the form
 \bea
 \Gamma[V^{++}]&=& \f{i}{4}\int d\zeta_{1}^{(-4)} du_1 V^{++}\,\int_0^1 dg \,
 \f{1}{\Delta}_1(\cD^+_1)^4(\cD^+_2)^4\f{\d^{14}(z_1-
 z_2)}{(u^+_1u^+_2)^3}\Big|_{2=1}\,. \label{Gamma}
 \eea

\subsection{Divergent part of the effective action}
In the framework of the proper-time technique, the inverse operator
$\f{1}{\Delta}$ is defined as follows
 \be
 \label{inverse}
 -\f{1}{\Delta}=\int_0^\infty d(is) e^{is\Delta}.
 \ee
To avoid the divergences on the intermediate steps it is necessary
to introduce a regularization. We will use a variant of dimensional
regularization (so called $\omega$ -regularization) accommodative
for regularization of the proper-time integral (see e.g. \cite{BK}).
The $\omega$-regularized version of the relation (\ref{inverse}) is
 \be
 \label{omegainverse}
 -(\f{1}{\Delta})_{reg}=\int_0^\infty
d(is)(is\mu^2)^{\omega} e^{is\Delta},
 \ee
where $\omega$ tends to zero after renormalization and $\mu$ is an
arbitrary parameter of mass dimension.

We will now concentrate on calculating the divergent part of the
effective action \eq{Gamma}. In the regularization scheme under
consideration, the divergences mean the pole terms of the form
$\f{1}{\omega}.$ Taking into account the relation
(\ref{omegainverse}) and the relations (\ref{GREEN}) one gets
 \be\label{omega}
\Gamma_{\rm div.part} = -\f{i}{4}\int
du_{1}d\zeta_{1}^{(-4)}V^{++}(1)\int_0^\infty
 d(is)(is\mu^2)^{\omega}
 e^{is\sB_1}(\cD_1^+)^4(\cD^+_2)^4\f{\delta^{14}(z_1-z_2)}{(u^+_1u^+_2)^3}\Big|^{2=1}_{\rm div.part}.
 \ee
In the expression \eq{omega} we used the explicit form of the
operator $\Delta$ \eq{sDelta} and omit the additional term
${\cW}^{-\a}\cD^+_\a$, thus it immediately coincide with the $\sB$
\eq{sBox}. We use the momentum representation of the delta function,
$\delta^{14}(z_1-z_2)=\delta^6(x_1-x_2)
\delta^4(\theta^{+}_1-\theta^{+}_2)\delta^4(\theta^{-}_1-\theta^{-}_2)$,
 \be\label{delta}
 {\bf 1}\d^{(14)}(z_1-z_2)=\int \f{d^6 p}{(2\pi)^6}
 e^{ip_a\rho^a} \d^{(8)}(\rho^{\a\pm})I(z,z')~,
 \ee
where
 \bea
 \rho^a=(x_1-x_2)^a -2i(\t_1^+-\t_2^+)\g^a\t^-_1, \q
 \rho^{\a\pm}=(\t^\pm_1-\t^\pm_2)^\a~.
 \eea
The parallel displacement propagator $I(z, z')$ is required only
beyond the one-loop approximation and not required for actual
one-loop calculations. Moving the exponential  to the left through
the differential operators,  $e^{is\sB}\d^{14}(z_1-z_2)$ becomes in
the coincidence limit
$\int\f{d^6p}{(2\pi)^6}e^{is\sB(X)}\d^8(\t_1-\t_1)|_{1=2}~.$ The
$X$’s being defined by $X_a=\cD_a+ip_a$ and
$X^-_\a=\cD^-_\a+2p_{\a\b}\rho^{-\b}.$ Note also that the term
$p_{\a\b}\rho^{-\b}$ in $X^-_\a$  vanishes in the coincidence limit
since there are no $\cD^+_\a$ operators  required to kill
$(\t^-_1-\t^-_2)$.

By expanding the $e^{is\stackrel{\frown}{\Box}_1(X)}$ in the
(\ref{omega}) and leaving only the terms relating to divergences one
gets
 \be\label{integral}
 e^{is\stackrel{\frown}{\Box}_1}(u_1^+u_2^+)(\cD_1^+)^4(\cD^-_1)^4\delta^{8}(\t_1-\t_2)|_{1=2}=
 -\int_0^\infty\f{d(is)}{(is)^3}(is\mu^2)^\omega e^{-ism^2}\{is
 Y^{++}+\f{(is)^2}{2}[\sB, Y^{++}]\}.
 \ee
Here we have introduced the infrared regulator $m^2$ which is a
natural element of the calculations of effective action in massless
gauge theories\footnote{The $4D, {\cal N}=2$ vector multiplet
contains the scalar which in principle can serve as the infrared
regulator. In $6D,\, \cN=(1,0)$ gauge theory such a scalar is
absent. However, such a scalar is present in $6D$ tensor hierarchy
(see e.g. \cite{BP15} and the references therein). Therefore one can
expect that in this case there will be no necessary to introduce the
special infrared regulator (see some attempt to study the effective
action in the hypermultiplet theory on background of tensor
hierarchy in \cite{BP15-1}).}. By calculating the proper-time
integral and extracting the pole terms one gets for the right hand
side of the above expression
 \be
 \f{1}{\omega}m^2Y^{++}-\f{1}{2\omega}\square
 Y^{++}-\f{1}{2\omega}W^{+\a}D^-_\a Y^{++}~.
 \ee
One can see that the divergent part of the effective action
\eq{omega} is proportional to $Y^{++}$ and it cancel if we assume
the 'on-shell' condition (classical equation of motion) $Y^{++}=0$
\eq{eqY} for the background fields. As a result the on-shell induced
effective action in $6D,\, \cN=(1,0)$ theory is ultraviolet finite.
It is a important distinctive property of six-dimensional
supersymmetric model.  In four dimensions the analogous
non-supersymmetric models and ${\cal N}=1$, ${\cal N}=2$
supersymmetric models are on-shell divergent.


\subsection{Low-energy effective action}

To find the complete low-energy effective Lagrangian we should fix
the appropriating background superfields. It was shown in
\cite{BP15}, \cite{BP15-1} that such a background is given by the
covariantly constant vector background without the auxiliary fields
('on-shell' background)
 \be\label{backgr}
  Y^{++} = 0\,,  \q D_a W^{\pm\,\a}=0\,,\q \cD^\pm_\a F_{ab}=0\,.
  \ee 
Thus, in the first equation of (\ref{backgr}) we assume that
background fields solve the classical equations of motion. Two other
equations in (\ref{backgr}) mean the covariant space-time
independence. For the 'on-shell' background under consideration the
operators $\Delta$ and ${\Omega^{--}}$ take a simple form and
depends only on the background fields $W^{+\alpha}$ and $D^-_\a
W^{+\b}$. Since the form of the effective Lagrangian is defined by
the coefficients of these operators we can conclude that the
low-energy effective Lagrangian should have the following general
form
 \be\label{eff.lagr_1} {\cal L}_{eff}^{(+4)}= {\cal L}_{eff}^{(+4)}(W^{+\alpha},
D^\mp_\a W^{\pm\b})~.
 \ee
The aim of this paper is to derive the complete low-energy effective
Lagrangian.

We use the proper time representation \eq{inverse} for the inverse
operator $ \Delta_1$  and act to the delta-function. It should be
noted that the operator $\Delta_1$ and the covariant derivative
${\cD}^+_1$ does not commute even for the background under
constrains (\ref{backgr}), indeed, $[\Delta, {\cD}^+_\a] = - N_\a^\b
{\cD}^+_\b $. This is a crucial differences of the theory under
consideration in comparison with $4D,\, {\cal N}=2$ hypermultiplet
theory where the corresponding operators commute on corresponding
on-shell background. Substituting (\ref{Delta}) into the effective
action (\ref{Gamma}) we obtain\footnote{Here we follow fourth paper
in \cite{KUZ}, where the analogous consideration in $4D, {\cal N}=2$
hypermultiplet theory has been carried out.}
 \bea
\Gamma[V^{++}] &=& - \f{i}{4}\int d\zeta_{1}^{(-4)} du_1 V^{++}\,
\int_0^1 dg \, \int_0^\infty d(is) \, e^{is \Delta_1}
 (\cD^+_1)^4(\cD^+_2)^4\f{\d^{14}(z_1-
 z_2)}{(u^+_1u^+_2)^3}\Big|_{2=1}  \nn \\
&=& - \f{i}{4}\int d\zeta_{1}^{(-4)} du_1 V^{++}\, \int_0^1 dg \,
\int_0^\infty d(is) \, (e^{-isN}\cD^+_1)^4
\Big\{(u^+_1u^+_2)(\cD^-_1)^4 \nn \\  &&
 \qquad -(u^-_1u^+_2)\Omega^{--}
  -4\sB\f{(u^-_1u^+_2)^2}{(u^+_1u^+_2)}\Big\}\,
e^{is \Delta_1} \d^{14}(z_1-z_2) \Big|_{2=1} \,. \label{L-1}
 \eea

The divergences of the effective action have been found in the
previous subsection. These divergences are proportional to $Y^{++}$,
therefore they vanish on the background under consideration. Hence
the effective Lagrangian is on-shell finite.

The terms in the braces in (\ref{L-1}) are considered as follows.
First term in the braces was analysed in previous subsection and it
was shown that it is proportional to $Y^{++}$ which is equal to zero
on the background under consideration. Last term in the braces
vanishes since there is no enough number of $D$-factors to
annihilate the delta-function of anticommuting variables with the
help of relation
 \be
 \label{identity2} (\cD^+)^4(\cD^-)^4\d^8(\t-\t')|_{\t=\t'}=1~.
 \ee
Second term in the braces in (\ref{L-1}) contains the operator
${\Omega}^{--}$ (\ref{O}). One can show that only first term in this
operator gives rise to the effective action. Indeed, the term $
D^{-}_{\alpha}W^{-\alpha}$ in this operator
is equal to $Y^{--}$  
which vanishes on the 'on-shell' background.

Taking into account the properties of $\Delta$ described, the
functional (\ref{L-1}) can be rewritten in the form:
 \bea\label{G}
 \Gamma[V^{++}]  &=& - \frac{i}{4} \int d\zeta_1^{(-4)}du_1 V^{++}
 \int_0^1 dg\,\Omega^{--} \,\int_0^\infty d(is)
(e^{-isN}\cD^+_1)^4 e^{is\Delta_1}\d^{(14)}(z_1-z_2)\bigg|_{2=1}
 \eea
A further simplification occurs in the case of a covariantly
constant vector multiplet (\ref{backgr}). Then, the first-order
operator appearing in $\Delta$,
 \be
 \label{Upsilon}
 \Upsilon = W^{+\a}\cD^-_\a - W^{-\a}\cD^+_\a
 \ee
turns out to commute with the vector covariant derivative $\cD_a$,
that allows us to represent $e^{is\Delta}$ in factorized form
$e^{is\Upsilon}e^{is\cD^a\cD_a}$. This allows us to build the
calculation of the heat kernel in the form
 \be
K(z_1,z_2|s)=e^{is\Upsilon}e^{is\cD^a\cD_a}\d^{14}(z_1-z_2)
 =e^{is\Upsilon}\tilde{K}(z_1,z_2|s).
 \ee
The reduced heat kernel $\tilde{K}(z_1,z_2|s)$ can now be evaluated
in the same way by generalizing the Schwinger construction
 \bea\label{sch}
 \tilde{K}(z_1,z_2|s)=\f{i}{(4\pi s)^3}\det{^{1/2}}\f{s F}
 {\sinh(s F)}e^{\f{i}{4}\rho^a(F\coth s F)_{ab}\rho^b}(\rho^+)^4 (\rho^-)^4 I(z_1,z_2)\,,
 \eea
where the determinant is computed with respect to the Lorentz
indices. To compute the kernel $K(z_1,z_2|s)$ we need to evaluate the
action of the $\exp{(is\Upsilon)}$ on the $\tilde{K}(z_1,z_2|s)$. The
formal result reads
 \bea
 \label{ker}
 K(z_1,z_2|s) = \f{i}{(4\pi is)^3}\det{}^{\f12}\bigg(\f{sF}{\sinh(sF)}\bigg)
e^{\f{i}{4} \r^a(s) (F\coth s F)_{ab} \r^b(s)}
\r^{+4}(s)\r^{-4}(s)I(z_1,z_2|s)\,,
 \eea
where we have denoted, $\r^A=(\r^a, \r^{\a+}, \r^{\a-})$,
 \bea
\r^A(s)=e^{is\U}\r^Ae^{-is\U}\,,  \quad I(z,z'|s) = e^{is\U}
I(z,z')\,.
 \eea
Using the formula $e^A B e^{-A} = B+[A,B]+\ldots$ and our
constraints on the background \eq{backgr} we
obtain\footnote{Here we use ${\cal D}^+_\a \r^{\b -} =\d_\a^\b$ and
${\cD}^-_\a \r^{\b+} =-\d_\a^\b$.}
 \bea
 \r^{\a+}(s) &=&\r^{\a+} - W^{\b + }{\cN}_\b^\a \,, \qquad \qquad
 \r^{\a-}(s) = \r^{\a-} - W^{\b - }{\cN}_\b^\a \,, \label{rho} \\
 \r^a(s) &=& \r^a - 2 \int_0^s dt\, W^{-}(t) \g^a \r^{+}(t)\,, \qquad
 W^{\a\,-}(s) =  W^{\b\,-}\,\big( e^{isN}\big)^\a_\b\,.
 \eea
Here we have used the notation \eqref{N} and ${\cal N}_\a^\b = \big(\f{e^{i s N}-1}{N} \big)_\a^\b$.
 We need not the explicit
expression for $I(z,z'|s)$ but it is easy to check by
differentiating over the proper time $s$ the identity
 \bea I(z_1,z_2|\,s)
 &=& \exp\left[\int_0^s dt\, \Sigma(z_1,z_2|\,t) \right] I(z_1,z_2)\,,
 \label{I(s)} \\
 \Sigma(z_1,z_2|\,t)&=& e^{it\U} \Sigma(z_1,z_2) e^{-it\U}\,,
 \label{Sigma(t)}
 \eea
and $\Sigma(z_1,z_2)$ is defined by
 \be
  (W^{+\a}\cD^-_\a  -
W^{-\a}\cD^+_\a)I(z_1,z_2)=\Sigma(z_1,z_2) I(z_1,z_2)\,.
\ee

Now let us return to the calculation of the effective action \eq{G}.
According to the \eq{backgr} we obtain for the $\Omega^{--}$
\eq{sDelta} in the \eq{G}
  \bea
\Gamma[V^{++}] &=&-\frac{i}{4}\int d\zeta_1^{(-4)}du_1 \,V^{++}
\int_0^\infty d(is)
 \nn \\ &&
\times \int_0^1 dg \Big\{ i\cD^{\a\b}\cD^-_\a\cD^-_\b +
4W^{-\a}\cD^-_\a \Big\}(e^{-isN}\cD^+_1)^4 K(z_1,z_2|s)\bigg|_{2=1}.
\label{G10}
 \eea
Let us consider the second term in the braces. Firstly we have to
note that $ \cD^-_\a (e^{-isN}\cD^+)^4 K(z_1,z_2|s)\big|_{2=1}\sim
(W^{+})^3$ and the connection \eq{identity} between $W^-_\a$ and
$W^+_\a$. Integrating by parts $D^{--}$ from $W^-_\a$  and using the
analyticity of the field $V^{++}$ and zero-curvature condition
\eq{zeroc} in Abelian case one can show that this term vanish.
Schematically it has the form
 \bea
 \int d\zeta^{(-4)}du V^{++}W^{-}
(W^+)^3\sim\int d\zeta^{(-4)}du V^{++}D^{--}(W^+)^4\sim \int
d\zeta^{(-4)}du V^{--} D^{++}(W^+)^4=0 \,. \nn
 \eea

Now we integrate by parts in the first term of the \eq{G10} keeping
in mind our restriction on the background \eq{backgr} (see the
similar analysis in the work \cite{BP15-1}). After that we have
  \bea \label{G1}
\Gamma[V^{++}] =-\frac{3i}{2}\int d\zeta_1^{(-4)}du_1 \, \int_0^1
dg\,\int_0^\infty d(is)\, W^{\a +}\,{\cD}^{-}_{1\,\a}
(e^{-isN}\cD^+_1)^4 K(z_1,z_2|s)\bigg|_{2=1}.
 \eea
Then we act by the operators ${\cD}^{-}_\a$ and $(e^{-isN}\cD^+)^4$
on the kernel \eqref{ker}. In the limit of coincidence, these
derivatives act only on the two-point function $\r^{+4}(s)$ and
$\r^{-4}(s)$. According to \eq{rho} and \eq{N} we have
 \bea
(e^{-isN}\cD^+_1)^4  \r^{-4}(s) \Big|_{2=1}&=& 1\,, \\
 W^{\a+}\,{\cD}^{-}_{1\,\a}\r^{+4}(s)\Big|_{2=1}&=&
 \f16 \ve_{\a\b\g\d}\,\ve^{\a'\b'\g'\d'}(W^+ e^{isN})_{\a'}^{\a} (W^+\cN )_{\b'}^\b (W^+\cN )_{\g'}^\g
 (W^+\cN)_{\d'}^\d \nn \\
 &=& (W^+)^4 \f{d}{d (is)} \Big((is)^4 \det \cN(is) \Big)\,.
 \eea

We recall that all gauge fields, except $V^{++}$ in the \eq{G},
linearly depend on the $g$. We make the change, $ isg \to s$, in the
proper time integral \eq{G1} and then integrate over $g$. As a
result we obtain
 \bea \label{G2}
 \Gamma[V^{++}] &=& \f{1}{(4\pi)^3}\int d \zeta^{(-4)}du \,(W^{+})^4\,
 \xi\Big(F,N\Big)\,, \\
 \xi(F,N) &=& \f{1}2 \int_0^{\infty}\f{ds}{s^{3}}\,e^{-s m^2 }\,
\f{d}{ds}\bigg( s^4 \det\Big(\f{e^{s N}-1}{sN}\Big)\bigg)
\det{}^{\f12}\bigg(\f{s F}{\sin{sF}}\bigg)\,.
 \label{xi}
 \eea
This is the final expression for complete low-energy effective
action in the theory under consideration\footnote{One of the basic
differences of the effective action in the $6D,\, \cN=(1,0)$
hypermultiplet theory with one in the $4D, {\cal N}=2$
hypermultiplet theory is the derivative with respect proper-time in
the integrand (\ref{G2}).}. One can show that the integrand in
(\ref{G2}) is analytic superfield under the on-shell condition
$Y^{++}=0$\footnote{See first of the identities (\ref{identity0}).}.
The effective action (\ref{G2}), (\ref{xi}) is manifestly gauge
invariant and manifestly $\cN=(1,0)$ supersymmetric by construction.
Expanding the effective action in power of the superspace strengths
and integrating over proper time, we will get the effective action
as an expansion in power of on-shell gauge and supersymmetric
invariants. In certain sense the effective action obtained can be
considered as the generating functional of on-shell\footnote{We
remind that the effective action is obtained under on-shell
condition $Y^{++}=0$.} $\cN=(1,0)$ supersymmetric invariants.

As an example to illustrate the general situation we derive several
such invariants on the base of the expressions \eq{G2}, (\ref{xi}).
Decomposition of the determinants in (\ref{xi}) up to the forth
order over $s$ has the form.
 \bea
\det{}^{\f12}\bigg(\f{sF}{\sin{sF}}\bigg) &=&
1+\f{s^2}{12}\tr F^2 + \f{s^4}{288} (\tr F^2)^2 + \f{s^4}{360} \tr F^4 + \ldots\,, \label{Det1}\\
\det\bigg(\f{e^{sN}-1}{sN}\bigg) &=& 1+\f{s^2}{24}\tr N^2 -
\f{s^4}{2880} \tr N^4 + \f{s^4}{1152}(\tr N^2)^2 +\ldots\,,
\label{Det2}
 \eea
where
 \bea
 \tr F^2 =  \f12 \tr N^2  \,, \qquad
  \tr F^4 = - \f14 \tr N^4 + \f{3}{16} (\tr N^2)^2
   \eea
Then we substitute \eq{Det1} and \eq{Det2} in to $\xi(F,N)$
 \bea
\xi = 2 \int_0^{\infty}ds\,e^{-s m^2 }\,\Big( 1+\f{5s^2}{48}\tr N^2
+\f{11 s^4}{1920} (\tr N^2)^2 -\f{s^4}{720} \tr N^4 +\ldots\Big)\,.
 \eea
Integrating over proper time $s$ and using the expression
 \bea
  \tr N^2 &=& D^+_\a D^+_\b W^{-\a}W^{-\b} \,, \\
   \tr N^4 &=& \f12\,(\tr N^2)^2 - 4 (D^{+})^4 (W^{-})^4
 \eea
we finally have
 \bea \label{G3}
 \Gamma[V^{++}] &=& \f{1}{32\pi^3 m^2}\int d \zeta^{(-4)}du \,(W^{+})^4\,
 \Big(1+\tfrac{5}{24\, m^4} D^+_\a D^+_\b W^{-\a}W^{-\b} \nn \\
 &&+\tfrac{29}{240\,m^8} (D^+_\a D^+_\b W^{-\a}W^{-\b})^2 - \tfrac{1}{15\,m^8} (D^{+})^4 (W^{-})^4
 +\ldots\Big)\,.
 \eea
This expression allows us to write down the following gauge and
$(1,0)$ supersymmetric invariants
 \bea
&& I_1 = (W^{+})^4\,, \qquad\qquad\qquad\qquad\qquad \,\,\, I_2 =
(W^{+})^4\, D^+_\a D^+_\b
 W^{-\a}W^{-\b} \,, \label{I12}\\
&& I_3 = (W^{+})^4\, (D^+_\a D^+_\b W^{-\a}W^{-\b})^2\,, \qquad
 I_4 = (W^{+})^4 (D^{+})^4 (W^{-})^4\,. \label{I34}
 \eea
The last invariant, $I_4$, corresponds to the term $(W^{+})^4
(W^{-})^4$ in the harmonic 6D, $\cN=(1,0)$ superspace in the central
basis or $\sim (W^i)^8$ in conventional one. It is clear that the
similar procedure allows us, in principle, to obtain any term in
expansion of the effective action in power series in superfield
strengths and their spinor derivatives.

It is useful to compare the invariants (\ref{I12}), (\ref{I34}) with
the on-shell invariants which were obtained in the recent paper
\cite{BIS} (see also \cite{S}). In the nomenclature of the work
\cite{BIS} the invariants (\ref{I12}), (\ref{I34}) have the
following canonical dimensions $d(I_{1})=8, d(I_{2})=12,
d(I_{3})=16, d(I_{4})=16$, where $d(I_{i})$ is the dimension of the
invariant $I_{i}, i=1,2,3,4.$\footnote{Following to the work
\cite{BIS} we suppose $d$ as a mass dimension of corresponding
component $6D$ Lagrangian. } The invariant $I_{1}$ coincides with
the corresponding $d=8$ invariant in the paper \cite{BIS}. The
similar invariant appeared also as a leading low-energy contribution
to effective action in the $6D$ model of hypermultiplet coupled to
tensor hierarchy \cite{BP15-1}. The invariants with the dimensions
$d=12, 16$ are not considered in \cite{BIS}. However, there is the
invariant of the dimension $d=10$ in \cite{BIS} which is absent in
our case. We will show that such an invariant vanishes in the case
of covariantly constant background (\ref{backgr}).

In analytic basis the $d=10$ invariant has the structure
\bea
 S^{(10)} = \int d\zeta^{-4}du\, \varepsilon_{\a\b\g\d}(D^+)^4
 (W^{+\a}W^{-\b}W^{+\g}W^{-\d})\,. \label{I10}
 \eea
However in our case of covariantly constant on-shell background
(\ref{backgr}), the $d=10$ invariant (\ref{I10}) equals to zero.
First of all we note that according to (\ref{identity0}) and first
condition in \eqref{backgr}
 \bea
 D^+_\a W^{+\b} =\f14\d^\a_\b Y^{++}=0\,.
 \eea
Now one uses constrains $D^{\pm}_\a F^{ab} = 0$, which is also a
part of background conditions (\ref{backgr}).  Hence
 \bea
 2 D^{\pm}_\a N_\g^\d = D^{\pm}_\a F_\g^\d =
(\g_{ab})_\g^{\d} D^{\pm}_\a F^{ab}= 0.
 \eea
As a result we have in the action \eqref{I10}
 \bea
(D^+)^4(W^{+\a}W^{-\b}W^{+\g}W^{-\d})\sim
\varepsilon^{\r\s\tau\kappa}D^+_\r\,D^+_\s (N_\tau^\b
N_\kappa^\d)W^{+\a}W^{+\g} = 0\,.
 \eea

Thus we see the reason why the $d=10$ invariant (\ref{I10}) is
absent in our approach. The background conditions (\ref{backgr})
contain not only on-shell condition $Y^{++}=0$ but also the
condition of space-time constancy of the background. Namely this
condition forbids $d=10$ invariant. It is worth emphasizing that the
last condition is crucial to obtain the closed Heisenberg-Euler type
effective action. However, if to set a goal to derive the on-shell
invariants from the effective action, we can relax the background
conditions (\ref{backgr}) eliminating the space-time constancy
conditions. In this case the effective action can be constructed in
form of expansion in derivatives of the superspace strengths and we
expect that all possible on-shell invariants will be obtained.


\section{Conclusion}
Let us briefly summarize the main results. We have considered a
problem of the induced effective action in the $6D,\, \cN=(1,0)$
hypermultiplet theory coupled to an external field of vector
multiplet. The theory is formulated in six dimensional (1,0)
harmonic superspace in terms of an unconstrained analytic
hypermultiplet superfield in the external superfield corresponding
to an Abelian vector multiplet. The effective action is computed in
the framework of superfield proper-time technique which allows us to
preserve a manifest gauge invariance and $\cN=(1,0)$ supersymmetry.
It was shown that the effective action under consideration is
on-shell finite.

To calculate the low-energy effective action it is sufficient to
consider a special background (\ref{backgr}). We have developed a
generic procedure for calculating the effective action on such a
background and found the complete effective action (\ref{G2}). As we
pointed out the divergences are absent on-shell (\ref{eqY}) and
therefore the complete low-energy effective action (\ref{G2}) is
automatically finite.

Note that the theory under consideration is anomalous \cite{ANOM}.
Therefore, in principle, the total effective action can contain,
besides above result, the additional effective action generating the
anomaly. Since the anomaly is stipulated by divergences and the
divergences are absent on the background under consideration, one
can expect that such an additional action is also absent in our
case. However, even if this additional action exists on the given
background, this will be independent contribution to the total
effective action (\ref{G2}) and its calculation is a separate
problem.

We expect that the obtained results have relation to the problem of
the effective action of a single isolated D5-brane \cite{Schwarz}.
However, to calculate the effective action for such a model we
should study a quantum vector/tensor + hypermultiplet system. Of
course, such a problem requires special consideration. Another
aspect, which is essential for finding the effective action of
D5-brane is that the calculations should be carried out on a
conformally broken phase of the $6D$ non-Abelian supersymmetric
gauge theory (see definition of this phase e.g. in \cite{SSWW}).
Nevertheless, we hope that the methods, developed in this paper, can
be used to analyze the general problem of the effective action of
the D5-brane.

The methods and results of the present work can be generalized in
the following directions: (i) calculation of the low-energy
effective action beyond the leading approximation, (ii) calculation
of the effective action in a non-Abelian theory in the broken phase,
(iii) calculation of the effective action of the quantum
vector/tensor+hypermultiplet system.

We announce with deep sorrow that our friend and co-author Nicolay
Pletnev passed away suddenly when the paper was practically
finalized.


\section*{Acknowledgments }

The authors are grateful to I.B. Samsonov for useful discussions.
The work was supported by Ministry Education and Science of Russian
Federation, project 2014/387/122. The authors are also grateful to
the RFBR grant, project No 15-02-03594, the RFBR-DFG grant, project
No 16-52-12012 and the LRSS grant, project No 6578.2016.2 for
partial support.

\bigskip

\end{document}